# Applicability of n-vicinity method for calculation of free energy of Ising model

**Boris Kryzhanovsky, Leonid Litinskii**

Center of Optical Neural Technologies,
Scientific Research Institute for System Analysis of Russian Academy of Sciences,
Nakhimov ave, 36-1, Moscow, 117218, Russia
*kryzhanov@mail.ru*        *litin@mail.ru*

## Abstract

In a previous work, the *n*-vicinity method for approximate calculation of the partition function of a spin system was proposed. The equation of state was obtained in the most general form. In the present paper, we analyze the applicability of this method for the Ising model on a *D*-dimensional cubic lattice. The equation of state is solved for an arbitrary dimension *D* and the behavior of the free energy is analyzed. As expected, for large dimensions, $D \geq 3$, the system demonstrates a phase transition of the second kind. In this case, we obtain an analytical expression for the critical value of the inverse temperature. When $3 \leq D \leq 7$ this expression is in a very good agreement with the results of computer simulations. In the case of small dimensions $D < 3$, there is a noticeable discrepancy with the known exact results.

## 1. Introduction

Nearly a century, the attention of researches was focused on the problem of calculation of the partition function in the Ising model. For some important and interesting models this problem can be solved accurately [1-4]. However, the existence of "difficult" cases for which no exact solutions have been found leaded to the development of methods solving this problem in the frameworks of different approximations. An extensive bibliography on this subject can be found in [5, 6]. Herein a numerous physical applications of the problem are also listed. Note, the problem is important not in physics only, but also in the problems of combinatorial optimization [7], the theory of machine learning [8] and computer processing of images [9], [10].

In [11-13] we proposed the *n*-vicinity method as a way of approximate calculation of the partition function for an arbitrary connection matrix. The main idea of our method is as follows. Let us fix the initial configuration (the initial state) $\mathbf{s}_0 \in \mathbf{R}^\mathbf{N}$ and define its *n*-vicinity $\Omega_n$ as the set of states that differ from $\mathbf{s}_0$ by the opposite values of *n* coordinates ($0 \leq n \leq N$). In a general case, the distribution of the energies of the states from the *n*-vicinity is unknown. However, in [12, 13] we found the exact expressions for the mean energy $E_n$ and the variance $\sigma_n^2$ in terms of the characteristics of the connection matrix. Numerous computer simulations show that the distribution of energies from $\Omega_n$ is a single-mode one and the Gaussian density with the calculated parameters $E_n$ and $\sigma_n^2$ approximates it rather accurately (see [13]). Then in the expression for the partition function $Z_N = \sum_{n=0}^{N} \sum_{\mathbf{s} \in \Omega_n} \exp(-\beta E(\mathbf{s}))$ the summation over the states from $\Omega_n$ can be replaced by the integration over the Gaussian measure, and in place of the summation over *n* we can pass to integration over the parameter $x = n/N$. After that, the partition function takes the form of the double integral $Z_N \approx \int dx \int dE \exp[-N \cdot F(x,E)]$ that can be calculated with the aid of the saddle-point method. The function $F(x,E)$ depends on the inverse temperature $\beta$, the external magnetic field $H$ (when it is present) and the characteristics of the connection matrix. In the general form the expression for $F(x,E)$ was obtained in [12, 13].

In the present publication, we describe an application of the *n*-vicinity method for analyzing the Ising model on a *D*-dimensional hypercube. The Ising model has been studied for almost hundred years, and a great variety of the obtained results can be used to verify our method and find the boundaries of its applicability. Below we list our results.

Assuming different interactions between the nearest-neighbors along different directions of the lattice, we obtain the equation of state (see Sec. 2, 3). It generalizes the Bragg-Williams equation. The effective coordination number *q* that defines the interaction of a spin with its nearest neighborhood is the main characteristic of the lattice. When the interactions with all the nearest-neighbors are the same, *q* is equal to the number of the nearest neighbors: $q = 2D$.

We solve the equation of state in Sec. 4 and Sec. 5. For the lattices of large dimensions $D > 3$, our solution reproduces all the known results for the Ising model. Namely, the presence of a phase transition of the second kind and the values of the critical exponents. We obtained an analytical expression for the critical value of the inverse temperature as function of $D$ (Sec. 4). It is in a good agreement with the results of computer simulations for the dimensions $3 \leq D \leq 7$ [14-17]. However, for the three-dimensional Ising model ($D = 3$) our approach does not provide the correct values of the critical exponents obtained with the aid of the renormalization group method [18, 19]. For small dimensions ($D < 3$) our approach is inapplicable. In the case of the one-dimensional Ising model, the same as in the mean field theory, we obtain an incorrect result indicating a phase transition at a finite temperature. For the two-dimensional Ising model ($D = 2$) our approach reproduces the known quantitative results rather well. However, contrary to the exact solution, it predicts a phase transition of the first kind (Sec. 5). In Sec. 6, we analyze the behavior of different physical characteristics (the magnetization, the internal energy and so on). In Sec. 7, we present our conclusions and discuss the obtained results. For details of calculations, see Appendixes.

## 2. Notations and general expressions

Let $\mathbf{T} = \left( T_{ij} \right)_1^N$ be a connection matrix corresponding to the $D$-dimensional Ising model with the nearest-neighbor interaction. The configuration $\mathbf{s}_0 = (1,1,...,1) \in \mathbf{R}^N$ is the ground state of the spin system and $\Omega_n$ is the $n$-vicinity of the ground state: $\Omega_n = \left\{ \mathbf{s} : \sum_{i=1}^N s_i = N - 2n \right\}, n = 0,.., N$. Our method of the partition function calculation is based on the assumption that the normal probability density accurately approximates the distribution of the energies from $\Omega_n$. In details this question was discussed in [13], where we also presented the exact expressions for the mean energy $E_n$ and the variance $\sigma_n^2$. Here we use their asymptotic forms only.

If $\mathbf{s} \in \Omega_n$ and $H$ is the uniform magnetic field, the energy per one spin is

$$E(\mathbf{s}, H) = -\frac{1}{2N} \sum_{i \neq j}^N T_{ij} s_i s_j - \frac{H}{N} \sum_{i=1}^N s_i = E(\mathbf{s}) - H \left( 1 - \frac{2n}{N} \right). \tag{1}$$

By $E_0$ we denote the energy of the ground state:

$$E_0 = E(\mathbf{s}_0) = -\frac{1}{2N} \sum_{i \neq j}^N T_{ij}. \tag{2}$$

Let $N \to \infty$ and let us introduce a parameter $x = n / N \in [0,1]$. In [13] we showed that the asymptotic expressions for the mean energy and the variance are:

$$E_x = \lim_{N \to \infty} E_n = E_0 (1 - 2x)^2, \quad \sigma_x^2 = \lim_{N \to \infty} \sigma_n^2 = \frac{8 \sum_{ij} T_{ij}^2}{N^2} x^2 (1-x)^2. \tag{3}$$

Appendix A proves that the asymptotic expression for the partition function has the form $Z_N \sim \int_0^1 dx \int_{E_0}^{|E_0|} \exp\left[ -N \cdot F(x, E) \right] dE$, where $N \gg 1$, and the function in the exponent is equal to

$$F(x, E) = L(x) + \beta \left[ E - H \cdot (1 - 2x) \right] + \frac{1}{2N} \left( \frac{E - E_x}{\sigma_x} \right)^2.$$

Here

$$L(x) = x \ln x + (1 - x) \ln(1 - x), \tag{4}$$

and $\beta = 1/T$ is the inverse temperature.

In the saddle-point method for calculation of the integral in the expression for $Z_N$, for each value of $\beta$ it is necessary to find the global minimum of the function $F(x, E)$. In other words, we have to solve the system of equations

$$\frac{\partial F}{\partial E} = \beta + \frac{E - E_x}{N\sigma_x^2} = 0, \quad \frac{\partial F}{\partial x} = \ln\frac{x}{1-x} + \frac{E - E_x}{N\sigma_x}\frac{\partial}{\partial x}\left(\frac{E - E_x}{\sigma_x}\right) + 2\beta H = 0.$$

From the first equation it follows that the minimum is reached inside the interval of the values of $x$, for which the inequality $\beta N\sigma_x^2 + E_0 - E_x \leq 0$ is fulfilled. Solving the first equation with regard to $\beta$ and substituting the result into the second equation, we see that the problem is reduced to minimization of the function of one variable

$$f(x) = L(x) + \beta E_x - \frac{\beta^2 N\sigma_x^2}{2} - \beta H(1 - 2x), \text{ when } \beta N\sigma_x^2 + E_0 - E_x \leq 0.$$

Using a suitable normalization, we introduce new notations

$$b = \beta\frac{\sum_{ij}T_{ij}^2}{\sum_{ij}T_{ij}} \quad, \quad q = \frac{\left(\sum_{ij}T_{ij}\right)^2}{N\sum_{ij}T_{ij}^2}, \ h = \frac{2\sum_{ij}T_{ij}}{\sum_{ij}T_{ij}^2}H. \tag{5}$$

In its final form the problem is as follows: for each value of "the normalized" inverse temperature $b$ it is necessary to find the global minimum of the function

$$f(x) = L(x) - \frac{qb}{2}\left[(1 - 2x)^2 + 8b(x(1-x))^2\right] - \frac{bh}{2}(1 - 2x) \tag{6}$$

at the interval $(0, x_b)$, where

$$x_b = \begin{cases} \dfrac{1}{2}, & \text{when } b \leq 1 \\ \dfrac{1 - \sqrt{1 - 1/b}}{2}, & \text{when } b \geq 1 \end{cases}. \tag{7}$$

The free energy as function of $b$ can be written as

$$\mathbf{f}(b) = b^{-1} \cdot \min_{x \in (0, x_b)} f(x). \tag{8}$$

Some notes are necessary. As far as $b \leq 1$, the function $f(x)$ defined in Eq. (6) has to be minimized at the interval $[0, 1/2]$. However, when $b$ is larger than 1, the right boundary of the interval $x_b$ depends on $b$ (see Eq.(7)). When $b$ tends to infinity, the right boundary tends to 0: $\lim_{b \to \infty} x_b = 0$.

For the $D$-dimensional Ising model with the interaction $J$ between the nearest-neighbors the equations (5) can be written as $\sum_{ij}T_{ij} = DNJ$, $\sum_{ij}T_{ij}^2 = DNJ^2/2$, $b = \beta J$, $q = 2D, h = 2H/J$. In this case, the parameter $q$ is equal to the number of the nearest-neighbors. However, if the interactions along different directions of the $D$-dimensional lattice are not the same, this parameter does not need to be an integer. For example, in the case of the two-dimensional Ising model with different interaction constants along the vertical and horizontal directions ($J$ and $K$, respectively) we have $q = 2\left(1 + \dfrac{2}{K/J + J/K}\right)$, and this parameter can take any value from the interval $(2, 4]$.

Similarly for the three-dimensional Ising model with different interaction constants $J, K, L$ we have $q = 2\dfrac{(J + K + L)^2}{J^2 + K^2 + L^2} \in (2, 6]$. In fact, the real number $q$ is the effective coordination number characterizing the interaction of a spin with its nearest neighbors.

Finally, we set $H = 0$. The results for a finite magnetic field will be the subject of a following paper.

# 3. Equation of state

*3.1. Main equation*

For each value of $b$ we find the minimum of the function

$$f(x) = L(x) - \frac{qb}{2}\left[(1-2x)^2 + 8b(x(1-x))^2\right] \tag{9}$$

on the interval $[0, x_b]$. For that purpose, we analyze the equation

$$\frac{\partial f}{\partial x} = \ln\frac{x}{1-x} + 2qb(1-2x)\left[1 - 4bx(1-x)\right] = 0 . \tag{10}$$

Evidently, $x_0 = 1/2$ is always the solution of equation (10): $x_0$ is called the trivial solution. To find other solutions we get rid of the trivial solution by transforming the equation (10) to the form

$$\frac{\ln\frac{1-x}{x}}{2(1-2x)} = qb\left[1 - 4bx(1-x)\right]. \tag{11}$$

Equation (11) is called the equation of state. By $l(x)$ we denote the left-hand side, and by $r(x,b)$ we denote the right-hand side of this equation, respectively:

$$l(x) = \frac{\ln\frac{1-x}{x}}{2(1-2x)}, \quad r(x,b) = qb\left[1 - 4bx(1-x)\right]. \tag{12}$$

The functions $l(x)$ and $r(x,b)$ are symmetrical with regard to the point $x_0 = 1/2$. To avoid the misunderstanding in what follows we examine the behavior of these functions on the interval $[0, 1/2]$ only.

The function $l(x)$ decreases monotonically from $+\infty$ at the left end of the interval to 1 at its right end: $l(x_0) = 1$. At the right end the values of the first and second derivatives of $l(x)$ are equal to $l'(x_0) = 0$ and $l''(x_0) = 8/3$, respectively. The quadratic polynomial $r(x,b)$ decreases monotonically from the value $r(0) = qb$ to the value $r(x_0) = qb(1-b)$ at the right end of the interval (see the lower panels of Fig.1). As far as the parameter $b$ increases, the inflection points or/and the local extrema of the function $f(x)$ appear (see the graphs at the upper panels of Fig.1). These transformations we analyze examining the relative position of the curves $l(x)$ and $r(x,b)$ (12).

It is clear that unless the value of $b$ is small, the curves $l(x)$ and $r(x,b)$ have no common points (see the lower panel of Fig.1a). In this case, the unique solution of the equation of state is the trivial solution $x_0 = 1/2$. The function $f(x)$ decreases monotonically from the value $f(0) = -qb/2$ at the left end of the interval to the value $f(x_0) = -\ln 2 - qb^2/4$ at its right end (Fig.1a), and the free energy (8) has the form:

$$\mathbf{f}(b) = \frac{f(x_0)}{b} = -\frac{\ln 2}{b} - \frac{qb}{4}, \quad b << 1 . \tag{13}$$

When the parameter $b$ increases, the curve $r(x,b)$ goes upward approaching the curve $l(x)$. For some value of $b$ the curves $l(x)$ and $r(x,b)$ touch each other. The x-coordinate of the tangency point can either be one of the inner points of the interval $(0, x_0)$ (see the lower panel of Fig.1b), or it can coincide with the right end of the interval and be equal to $x_0 = 0.5$ (see the lower panel of Fig.1c).

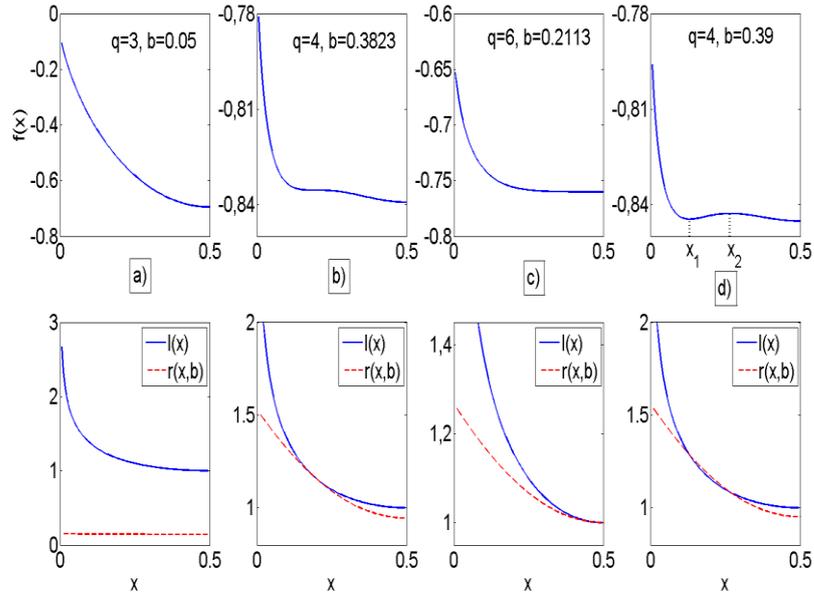

Fig. 1. Graphs of the function $f(x)$ (upper panels) and the curves $l(x)$ and $r(x,b)$ (lower panels) for different values of $q$ and $b$.

The tangency of the curves $l(x)$ and $r(x,b)$ at the inner point of the interval means that the inflection point of the function $f(x)$ appears. At this point the derivative of the function $f(x)$ is equal to zero (see the upper panel of Fig.1b), but since it is not the minimum of the function $f(x)$ this solution of equation (11) is not of interest for us. However, when the parameter $b$ increases, two extrema of the function $f(x)$ appear in place of the inflection point. They are the local minimum $x_1$ and the local maximum $x_2$ (compare the upper panels of Fig. 1b and Fig. 1d). The new local minimum is a nontrivial solution of the equation (10). Its depth is to be compared with the minimum at the point $x_0 = 1/2$. Let us find out when the curves $l(x)$ and $r(x,b)$ touch each other at the inner point of the interval $(0, x_0)$.

### 3.2. The tangency of the curves $l(x)$ and $r(x,b)$ at the inner point

This case is described by the system of equations

$$\begin{cases} l(x) = r(x,b) \\ l'(x) = r'(x,b) \neq 0 \end{cases}.$$  (14)

The nonequality sign in the second equation excludes the tangency of the curves at the right end of the interval at the point $x_0 = 1/2$. We solve the system of equations (14) by the variable separation (see Appendix B). Here we only formulate the obtained result.

**Statement 1.** The curves $l(x)$ and $r(x,b)$ touch each other at the inner point of the interval $(0, x_0)$ only if $q < 16/3$. The tangency of the curves takes place at the point $x_t$ (the subscript "t" indicates the tangency of the curves), which is the solution of the equation

$$G(x) = \frac{\left[\xi(x) + \varphi(x)\right]^2}{\xi(x)} = q;$$

the expressions for the functions $\varphi(x)$ and $\xi(x)$ are given in Appendix B (see (B1)). The solid line in Fig.2 shows the graph of the function $G(x)$. It increases monotonically from 0 to $G(1/2) = 16/3$. The tangency point $x_t$ corresponds to the intersection of the graph of the function $G(x)$ with the straight line parallel to the abscissa axis at the distance $q$ from it (see Fig.2). The value $b_t$ of the parameter $b$ under which the tangency of the curves $l(x)$ and $r(x,b)$ takes place can be calculated using the obtained value of $x_t$:

$$b_t = B(x_t), \text{ where } B(x) = \frac{\xi(x)}{\xi(x) + \varphi(x)}. \qquad (15)$$

In Fig.2 the dashed line is the graph of the function $B(x)$. It decreases monotonically from $+\infty$ at the left end of the interval to its minimal value $B(1/2) = 0.25$.

Note that Fig.2 can also be used for the graphical solution of a different problem. Namely, for a given value $b_t > 0.25$ we, first, can define the tangency point $x_t$ of the curves $l(x)$ and $r(x,b)$, and then we can calculate the value of $q$ for which the curves $l(x)$ and $r(x,b)$ touch each other at the point $x_t$ when $b = b_t$.

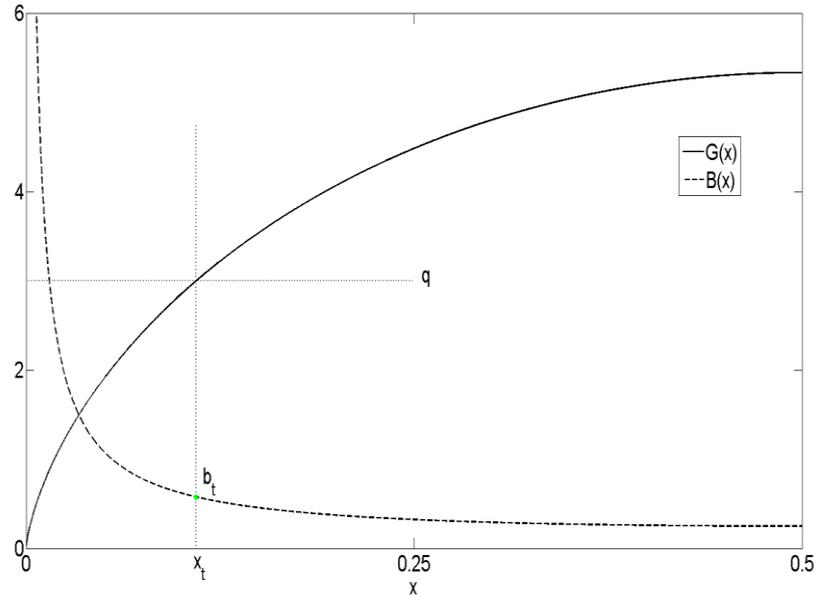

Fig.2. Functions $G(x)$ and $B(x)$ used for determination of the tangency point $x_t$ of the curves $l(x)$ and $r(x,b)$ at the inner point of the interval $(0, x_0)$.

## 4. Interval $q \geq 16/3$

### 4.1. The critical temperature

Let us analyze what happens when $q \geq 16/3$. In this case the curves $l(x)$ and $r(x,b)$ (12) never have a tangency point inside the interval $(0, x_0)$. Only the touching of the curves $l(x)$ and $r(x,b)$ at the right end of the interval at the point $x_0 = 0.5$ is possible (see Fig.1c). In this case the tangency condition is $l(x_0) = r(x_0, b)$, and it results in the quadratic equation $1 = qb(1-b)$. This equation has two roots

$$b_{\pm} = \frac{1 \pm \sqrt{1 - 4/q}}{2}. \qquad (16)$$

At the point $x_0 = 1/2$ the curves $l(x)$ and $r(x,b)$ have the tangency points at two values of the parameter $b$: when $b = b_-$ and when $b = b_+$. The first tangency occurs when $b$ becomes equal to the least of the two roots (16). Let us denote it as $b_c$:

$$b_c = b_- = \frac{1 - \sqrt{1 - 4/q}}{2}. \qquad (17)$$

As soon as $b$ exceeds $b_c$, $x_0$ becomes the maximum point of the function $f(x)$:

$$f''(x_0) \sim 1 - qb(1-b) < 0, \text{ when } b_- < b < b_+. \qquad (18)$$

In the same time near the point $x_0$ (to the left of it) a new minimum point of the function $f(x)$ appears. We denote it as $x_1$: $x_1 < x_0$ - see the upper panel of Fig.3a.

With further increase of $b$ the minimum point $x_1$ shifts toward 0 and the depth of the minimum increases. As before, the curve $r(x,b)$ moves upward. When the value of $b$ becomes equal to 0.5, the right end of the curve $r(x,b)$ is the highest: $r(x_0,1/2) = \max_b r(x_0,b) = q/4$ (Fig.3b). After that, the right end of curve $r(x,b)$ starts going down, and the left end $r(0,b)$ rises. When $b$ becomes equal to the second root of the equation (16) $b = b_+$, the curves $l(x)$ and $r(x,b)$ again touch each other at the point $x_0 = 0.5$. As soon as $b$ exceeds the value $b_+$, the point $x_0$ again becomes the minimum point of the function $f(x)$ (see Eq.(18)). The maximum of the function $f(x)$ appears to the left of $x_0$, at the point $x_2$ (see Fig.3c). Now the function $f(x)$ has two minima. They are the global minimum at the point $x_1$ and the local one at the point $x_0$. Under further increase of $b$ up to the value $b = 1$ the point of the global minimum $x_1$ shifts steadily to 0 and the minimum deepens. The estimate shows that when $b = 1$ the difference between the depths of the global and the local minima is $f(x_1) - f(x_0)|_{b=1} < \ln 2 - q/4 < 0$.

As soon as $b$ becomes larger than 1, the right boundary of the interval where $x$ changes becomes a variable (see Eq.(7) for $x_b$). In this case, we have to compare the global minimum $f(x_1)$ not with $f(x_0)$ but with the much greater value $f(x_b)$ (see Fig. 3b). When $b \to \infty$ the interval $(0, x_b)$ shrinks to the origin and the global minimum point tends to 0. This result is correct: when the temperature tends to zero the spin system tends to the ground state $\mathbf{s}_0 = (1,1,...,1)$.

### 4.2. Phase transition of the II kind

As far as the parameter $b$ does not exceed $b_c$ (see Eq.(17)), the free energy has the form (13). If $b$ becomes larger than $b_c$, the free energy has to be calculated by substituting $x_1$ in Eq.(9). When $b = b_c$, the system exhibits a phase transition of the second kind.

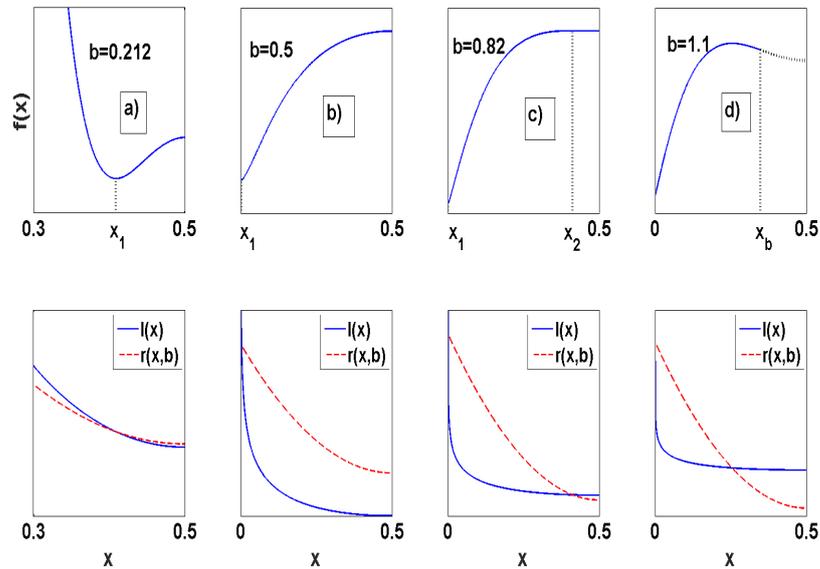

Fig.3. The Function $f(x)$ (upper panels) and the curves $l(x)$, $r(x,b)$ (lower panels) for $q = 6$ and different values of $b$ – see the body of the paper.

Let us substitute the values $q = 6, 8, 10, 12, 14$ into equation (17) and calculate the critical values of the inverse temperature for the $3D$-, $4D$-, $5D$-, $6D$- and $7D$- Ising models. We show the obtained numbers in the second row of

Table 1. The third row contains the critical values of the inverse temperatures obtained by means of computer simulations [14 − 17]. Our theoretical estimates are in a good agreement with these results.

**Table 1**. Critical values of the inverse temperature $b_c$ for the $D$-dimensional Ising models

| Dimension of lattice | 3D | 4D | 5D | 6D | 7D |
|---|---|---|---|---|---|
| Our theory: Eq. (17) | 0.2113 | 0.1464 | 0.1127 | 0.0918 | 0.0774 |
| Computer simulations | 0.2216 | 0.1489 | 0.1139 | 0.0923 | 0.0777 |

In Fig.4, we present the data from Table 1 in the graphic form. The solid line is the result of computer simulations, the dashed line is our estimate according Eq.(17), the dotted line is the estimate $b_c = 1/2D$ of the mean field theory (the Bragg-Williams equation [1]). We see that our formula describes the computer simulations much better. The larger $D$ the better the agreement between our theory and computer simulations. Recall the parameter $q$ can be an arbitrary one and it can take not only even integer values (see the note in the end of Sec. 2).

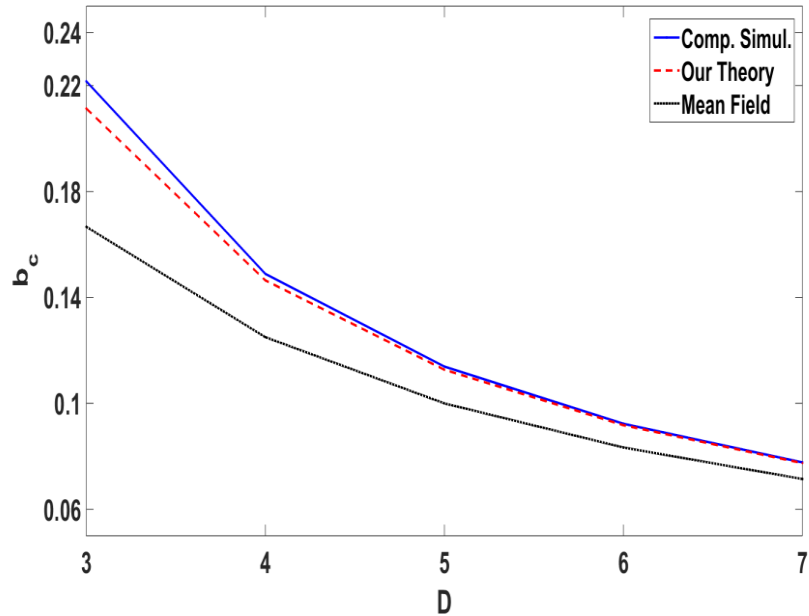

Fig.4. Dependence of the critical temperature $b_c$ on the dimension of the lattice $D$: computer simulations (solid line), Eq.(17) (dashed line), the mean field model (dotted line).

The critical exponents for $q \geq 16/3$ can be obtained by carrying out the standard calculations [1]. For all $q$ they have the classical values: $\alpha = 0$, $\beta = 1/2$, $\gamma = \gamma' = 1$, $\delta = 3$. For $D > 3$ the same values were obtained by other authors [21, 22]. For $D = 3$ the generally accepted are not the classical values of the critical exponents, but the ones calculated by means of the renormalization group method [18, 19].

# 5. Interval $q < 16/3$

## 5.1. The local minimum

For each $q$ less than 16/3 the function $f(x)$ has two extrema for the values of $b$ larger than $b_t$ (15): the minimum at the point $x_1$ and the maximum at the point $x_2$: $x_1 < x_2$ (see the upper panel of Fig.1d). When $b$ increases, the minimum point $x_1$ shifts to the origin, and the maximum point tends to the right end of the interval that is to the point $x_0$. When $f(x_1)$ becomes equal to $f(x_0)$ the global minimum jumps from the point $x_0$ into the point $x_1$.

This situation is described by the system of equations:

$$\begin{aligned} l(x) &= r(x,b) \\ f(x) &= f(x_0) \end{aligned}, \quad b \le 1. \tag{19}$$

The condition $b \le 1$ means that we compare the inner local minimum $f(x_1)$ with $f(x_0)$ but not with $f(x_b)$, which is the value of the function $f(x)$ at the varying boundary $x_b$. We solve the system of equations (19) in the same way as the system of equations (14) (see Appendix C). In this case, the main result is as follows.

**Statement 2.** There is such critical value $q_c$ of the parameter $q$,

$$q_c \approx 2.75,$$

that for $q$ less than $q_c$ ($q < q_c$) the value of $f(x_1)$ will never be less than the depth $f(x_0)$ of the minimum at the point $x_0$. For any $q$ from the interval $q_c \le q < 16/3$ there is a unique value $b_j$ that is less than 1, for which the local minimum at the point $x_1(b)$ coincides with the depth of the global minimum at the point $x_0$: $f(x_1(b_j)) = f(x_0)$. When $b$ becomes equal to $b_j$ the global minimum of the function $f(x)$ jumps from the point $x_0$ to the point $x_j = x_1(b_j)$ (the subscript "j" stands for "jump").

The point $x_j$ is defined as the intersection of the straight line parallel to the abscissa axis at a distance $q$ from it and the graph of the function $G_1(x)$:

$$G_1(x) = \frac{\left[\xi_1(x) + \varphi_1(x)\right]^2}{\xi_1(x)} = q. \tag{20}$$

The functions $\varphi_1(x)$ and $\xi_1(x)$ entering Eq.(20) are defined in Eq. (C2). In Fig.5 the solid line corresponds to the graph of the function $G_1(x)$. The value $b_j$ is calculated using the value of the function $B_1(x)$ at the point $x_j$:

$$b_j = B_1(x_j), \text{ where } B_1(x) = \frac{\xi_1(x)}{\xi_1(x) + \varphi_1(x)}.$$

In Fig.5 the dashed line shows the graph of the function $B_1(x)$.

It can be shown that

$$\lim_{x \to 0} B_1(x) = 2, \quad \lim_{x \to 0} G_1(x) \approx -\frac{\ln x}{4} \to +\infty, \quad \lim_{x \to 0.5} B_1(x) = 0.25, \quad \lim_{x \to 0.5} G_1(x) = 16/3.$$

When $x \to 0$, the functions $G_1(x)$ and $B_1(x)$ converge logarithmically slow to their limiting values. When $x$ increases, the function $B_1(x)$ decreases monotonically from $B_1(0) = 2$ to $B_1(1/2) = 0.25$. The function $G_1(x)$ is nonmonotonic. At first, this function decreases from $+\infty$ to its minimal value $\min G_1(x) \approx 2.75$ and at that, the function $B_1(x)$ decreases from 2 to 1. After passing through the minimum, the function $G_1(x)$ increases monotonically and tends to the value $G_1(1/2) = 16/3$ at the right end of the interval.

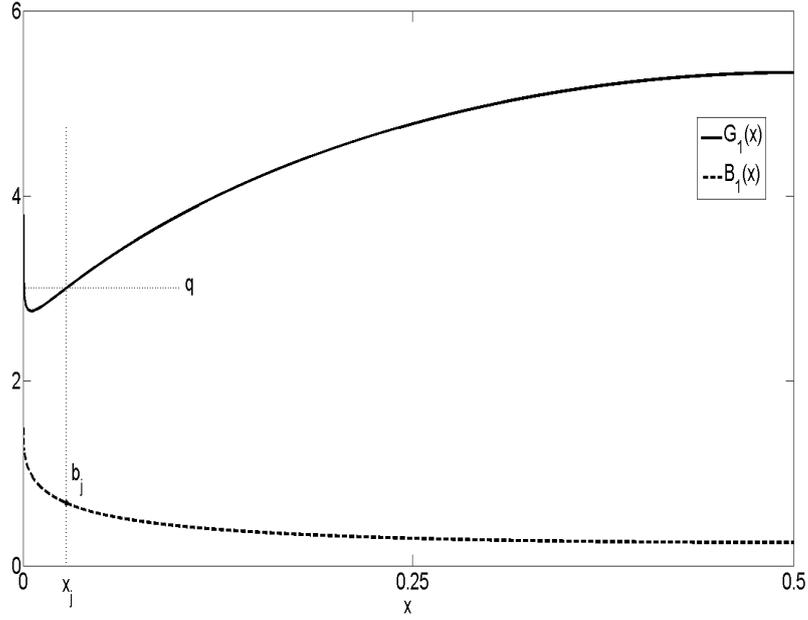

Fig.5. The graphs of functions $G_1(x)$ and $B_1(x)$ for determination of the jump point of the global minimum.

From Fig.5 we see that Eq. (20) has no solutions when $q < q_c \approx 2.75$. Consequently, for these $q$ the local minimum $f(x_1)$ never becomes deeper than $f(x_0)$. On the contrary, if $q_c \leq q < 16/3$ Eq. (20) has exactly two solutions (recall that when $x \to 0$ the function $G_1(x)$ logarithmically slow tends to infinity). From the two solutions, we have to choose the larger one, since the corresponding value of the parameter $b_j = B_1(x_j)$ is less than one. Now we use the obtained results to analyze the behavior of the spin system.

### 5.2. Interval $q < q_c$

For such values of $q$ the local minimum at the point $x_1$ never becomes deeper than the minimum at the point $x_0$. Consequently, as far as $b$ remains less than 1, the global minimum of the function $f(x)$ is at the point $x_0$. When $b$ exceeds 1, the right boundary $x_b$ of the interval becomes a variable (see Eq.(7)), and the global minimum of the function $f(x)$ shifts to the point $x_b$. The expression for the free energy has the form

$$\mathbf{f}(b) = b^{-1} \begin{cases} -\ln 2 - \dfrac{qb^2}{4}, & b \leq 1 \\ L(x_b) - \dfrac{q}{4}(2b-1), & b \geq 1. \end{cases}$$

When $b = 1$ the first derivative of the free energy has a jump,

$$\left.\frac{d\mathbf{f}}{db}\right|_{b=1-0} = -\frac{q}{2}, \quad \left.\frac{d\mathbf{f}}{db}\right|_{b=1+0} = \frac{1-q}{2},$$

which is the evidence of a phase transition of the first kind. This is true for any value of $q < q_c$, particularly, for $q = 2$ that corresponds to the one dimensional Ising model. On the other hand, the exact solution for the one dimensional Ising model [1] does not show any phase transition at finite temperatures. We admit that in the region $q < q_c$ our method leads to incorrect results.

### 5.3. Interval $q_c \leq q < 16/3$

For such $q$ the depth of the inner local minimum becomes equal to the depth of the global minimum at the point $x_0$ at a certain value $b_j$ ($b_j < 1$). At the same time the global minimum of the function $f(x)$ jumps from the point $x_0$ to the point $x_j = x_1(b_j)$. The magnetization of the state has a jump discontinuity. It changes from $m_0 = 0$ to $m_j = 1 - 2x_j$, and this is the evidence of a phase transition of the first kind. With further increase of $b$ ($b \to \infty$) the global minimum point $x_1(b)$ steadily tends to 0 and the system tends to the ground state.

The value $q = 4$ corresponding to the $2D$ Ising model belongs to this interval. For the $2D$ Ising model, our approach predicts a phase transition of the first kind at $b_j \approx 0.3912$. However the exact Onsager solution shows a phase transition of the second kind at larger critical value $b_c \approx 0.4407$ [1]. For $q$ from this interval the results of our method are ambiguous (see the next section).

## 6. Physical characteristics

Our calculations of different physical characteristics confirms the performed analysis of the equation of state (11). In Fig. 6 the dependence of the magnetization $m = 1 - 2x$ on $b$ is shown for different values of $q$. We have seen above that as far as $b$ is not large enough, the global minimum is at the point $x_0 = 1/2$ and the magnetization is equal to 0. When $b$ becomes equal to the critical value, the global minimum leaves the point $x_0$ and the magnetization becomes nonzero. For $q = 6$ the critical value is $b = b_c$ (17), and the magnetization changes without the jump discontinuity. For other $q$ the critical value is $b = b_j$, the global minimum jumps from the point $x_0$ to the minimum point $x_j$ (see above). When $q < 16/3$ the graphs clearly show the jumps of the magnetization; it changes from zero to $m_j = 1 - 2x_j$.

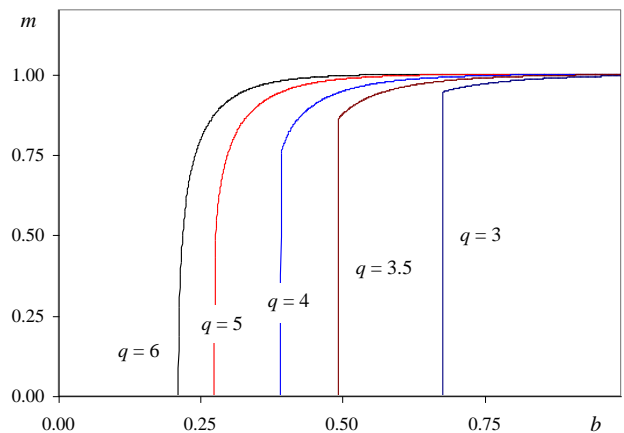

Fig. 6. The magnetization $m$ as function of $b$ for different values of $q$.

The internal energy behaves similarly. In Fig. 7 the dependencies of the internal energy normalized to the ground state energy on $b$ are shown. As it has to be, at first for each $q$ the internal energy depends on $b$ linearly. The dependence changes when $b$ exceeds the corresponding critical value. For $q = 6$, the function $U$ decreases continuously up to -1. For $q \in [q_c, 16/3)$ the internal energy $U$ has a jump discontinuity at the critical point, and then the system tends to the ground state.

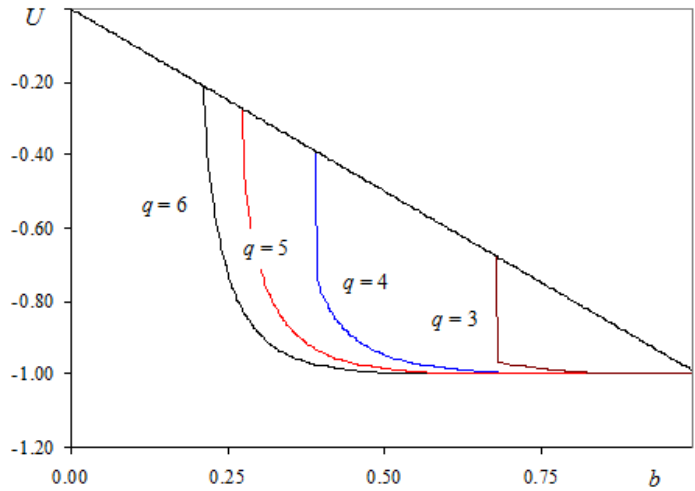

Fig. 7. Normalized internal energy $U$ for different values of $q$.

Let us now discuss the results for the 2D Ising model. In this case, our approach predicts the jump discontinuity of the magnetization. This error is because the normal density approximates accurately enough only the central part of the energy distribution. The tails of the distribution differ from the normal density [13]. From the analysis of the equation of state, we see that by a comparatively small correction of the curves $l(x)$ and $r(x,b)$ (12) the jump discontinuity of the magnetization can be eliminated. We suppose that the correction can be done by using a more general method based on the Edgeworth expansion [20] in place of the normal approximation. This is the subject of further analysis. If the jump discontinuity of the magnetization is ignored and one is interested in the behavior of the free energy only, our results for the 2D Ising model are quite decent.

In Fig. 8 for the 2D Ising model we show the dependence of the critical value of the inverse temperature $\beta_c$ on the ratio $K$ of the interactions between spins along the vertical and horizontal directions. We see that overall our results correlate with the exact solution.

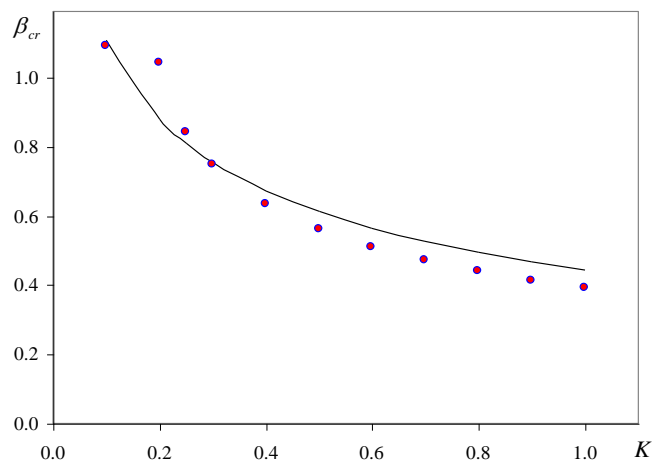

Fig. 8. The dependence of the critical temperature on the ratio of interactions $K$ for the 2D Ising model: the solid line is the Onsager solution; the circles are our results.

In Fig. 9 the graphs of the free energy $\mathbf{f}(\beta)$ for different values of $K$ are shown. The solid lines correspond to the exact Onsager solution; the circles are the results of our approach. It is clearly seen that the difference between the exact theory and our approximation becomes noticeable only in the vicinity of the critical value $\beta_c$. To make this statement more demonstrative, in Fig. 10 we show the graphs of the relative error $\left(\mathbf{f}_{our}(\beta) - \mathbf{f}_{Ons}(\beta)\right)/\mathbf{f}_{Ons}(\beta)$. The subscripts "Ons" and "our" indicate the Onsager and our solutions, respectively. The maximal relative error of the order of 1% corresponds to the critical value $\beta_c$.

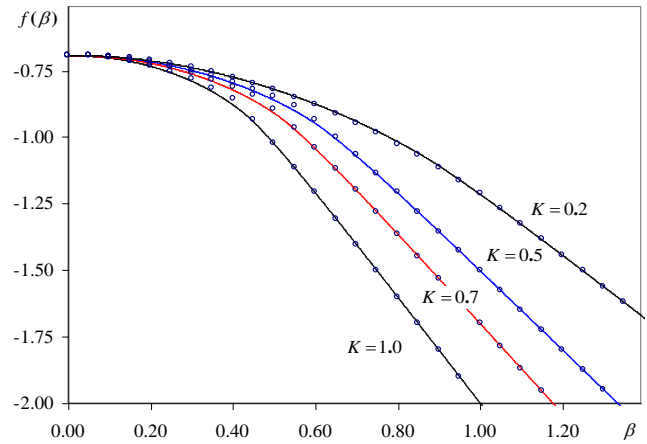

Fig.9. 2$D$ Ising model: free energy $\mathbf{f}(\beta)$ for $K = 0.2, 0.5, 0.7, 1.0$. Solid lines are the Onsager theory; circles are our results.

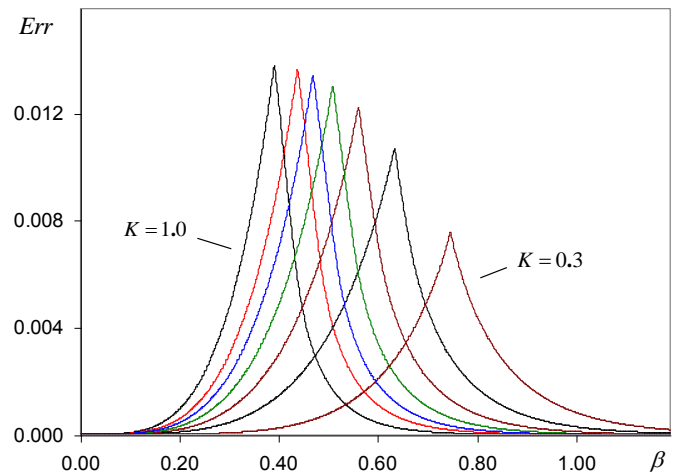

Fig.10. 2$D$ Ising model: relative error for different values of $K$. From left to right: $K = 1, 0.8, 0.7, 0.6, 0.5, 0.4, 0.3$.

## 7. Discussion and conclusions

Let us make the approximation used in our approach more accurate. In its framework, when going from the summation over the $n$-vicinity $\Omega_n$ to integration over energy, we replace the true but unknown energy distribution $P_n(E)$ by the Gaussian density with the known mean $E_x$ and the variance $\sigma_x^2$ (see Introduction). Computer simulations show that the normal distribution approximates very well the central part of $P_n(E)$ [13]. At that, the tails of the distribution $P_n(E)$ are not always Gaussian (the "fat tails"). Just this is the source of the error. Let us show that the error decreases when the coordination number $q$ increases.

In Eq.(A2) we pass to dimensionless variables $\varepsilon = E/|E_0|$, $\varepsilon_x = E_x/|E_0|$, $\tilde{\sigma}_x = \sigma_x/|E_0|$ and $\bar{\beta} = \hat{\beta}|E_0|$. Then Eq.(A2) takes the form

$$\frac{1}{C_N^n} \sum_{s \in \Omega_n} e^{-\hat{\beta}E(s)} \approx \frac{1}{\sqrt{2\pi}\tilde{\sigma}_x} \int_{-1}^{1} \exp\left[-\bar{\beta}\varepsilon - \frac{1}{2}\left(\frac{\varepsilon - \varepsilon_x}{\tilde{\sigma}_x}\right)^2\right] d\varepsilon. \qquad (21)$$

In this expression the limits of integration do not depends on the parameters of the problem. The accuracy of the equality (21) depends on the value of the variance $\tilde{\sigma}_x$: the less $\tilde{\sigma}_x$ the less the contribution of the tails in the integral in the right-hand side of Eq.(21). For the Ising model we have

$$\tilde{\sigma}_x = 4x(1-x)\sqrt{\frac{2}{qN}}.$$

As we see the value of $\tilde{\sigma}_x$ decreases when $q$ increases. Consequently, the accuracy of the estimate of the integral (21) also increase as well as the accuracy of our approximation.

For cubic lattices with large value of the coordination number $q > 16/3$, the results of the $n$-vicinity method are in a good agreement with the previously obtained results of other authors. Our method predicts a phase transition of the second kind, and the analytical expression (17) describes accurately the results of computer simulations. For all $q$ from this region the critical exponents are equal to their classical values. For $D \geq 4$ the same was proved in [21], [22] by means of more rigorous methods. For $D = 3$ the renormalization group method provides the critical exponents that are not equal to the classical ones.

The results for the lattices with small coordination numbers $q_c < q < 16/3$ have to be treated critically. For example, in the case of the $2D$ Ising model, the $n$-vicinity method predicts the jump discontinuity of the magnetization, and that contradicts to the known exact solution of this problem. Moreover, when $q < q_c \approx 2.75$ the $n$-vicinity method is not at all applicable.

From here, the boundaries of applicability of the $n$-vicinity method are clearly seen. One can reliably use this method for large dimensions $D \geq 3$ having in mind that for $D = 3$ the critical exponents mismatch the classical ones. For $D = 2$ we can calculate the dependence of the free energy on the inverse temperature. Near the critical temperature, it differs from the accurate one by 1-2 percent. Note, in the modern methods of the scene analysis and the computer data processing [23-26] the normalization constant has to be repeatedly calculated. It turns out to be equal to the partition function for fixed values of the external parameters. Here the method of quick calculation of the partition function (even an approximate and not too accurate one) can be useful. In this case, nobody cares about the value of the critical temperature.

The above-said is related to the Ising model with the interaction between the nearest neighbors. When the interaction between the next neighbors or even the more distant neighbors is taken into account, it results in the increase of the effective coordination number $q$. Almost certainly, it will make better the agreement between the results of the $n$-vicinity method and the exact solutions or computer simulations.

Our approach can be improved if in place of the unknown function $P_n(E)$ we use a more general expression than the normal density. For this, the Edgeworth expansion is a suitable tool [20]. In this case, we expand $P_n(E)$ into a series of functions $\dfrac{H_k(x)}{\sqrt{2\pi}} e^{-\frac{x^2}{2}}$, where $H_k(x)$ is the Hermitian polynomial, $k = 0,1,2...$. Note the approximation of $P_n(E)$ by the normal distribution is equivalent to using of the first term of the Edgeworth expansion only. The main problem is to calculate the higher moments of distribution $P_n(E)$ the same as it has been done in [12,13] when calculating the mean energy $E_n$ and the variance $\sigma_n^2$.

## Acknowledgments

The work was supported by the Russian Basic Research Foundation (grants №15-07-04861 and 16-01-00626). The authors are grateful to Dr. Inna Kaganova, Dr. Iakov Karandashev, and Dr. Magomed Malsagov for stimulating discussions and their help during the course of the work.

## Appendix A

Let the ground state of the Ising model $\mathbf{s}_0 = (1,1,1,...,1)$ be the initial configuration. The n-vicinity $\Omega_n$ of $\mathbf{s}_0$ consists of configurations with the same values of the magnetization

$$m = \frac{1}{N}\sum_{i=1}^{N} s_i = 1 - \frac{2n}{N}, \; n = 0,1,...,N \; .$$

We use the expression (1) for the energy per spin and introduce $\hat{\beta} = \beta N$, where $\beta = 1/T$ is the inverse temperature. The partition function can be written as

$$Z_N = \sum_{\mathbf{s}} e^{-\hat{\beta}E(\mathbf{s},H)} = \sum_{n=0}^{N} e^{\hat{\beta}H\left(1-\frac{2n}{N}\right)} \sum_{\mathbf{s}\in\Omega_n} e^{-\hat{\beta}E(\mathbf{s})} = \sum_{n=0}^{N} C_N^n e^{\hat{\beta}H\left(1-\frac{2n}{N}\right)} \left(\frac{1}{C_N^n}\sum_{\mathbf{s}\in\Omega_n} e^{-\hat{\beta}E(\mathbf{s})}\right). \tag{A1}$$

In Eq.(A1) we replace the summation by integration supposing that $N \to \infty$. For that, we introduce the real variable $x = n/N$ and use the Stirling formula $\binom{N}{xN} \sim \exp(-N \cdot L(x))$, where $L(x)$ is given by Eq.(4). The summation over $n$ in Eq.(A1) is replaced by integration according to the rule $\sum_{n=0}^{N} \to \int_0^1 dx$. Integration over the normal distribution with the known mean value $E_x$ and the variance $\sigma_x^2$ replaces the averaging over $\Omega_n$:

$$\frac{1}{C_N^n}\sum_{\mathbf{s}\in\Omega_n} e^{-\hat{\beta}E(\mathbf{s})} \approx \frac{1}{\sqrt{2\pi}\sigma_x} \int_{E_0}^{|E_0|} \exp\left[-\hat{\beta}E - \frac{1}{2}\left(\frac{E-E_x}{\sigma_x}\right)^2\right] dE \; . \tag{A2}$$

Then

$$Z_N \sim \int_0^1 \exp\left(-NL(x) + \hat{\beta}H(1-2x)\right) dx \int_{E_0}^{|E_0|} \exp\left(-\hat{\beta}E - \frac{N}{2}\left(\frac{E-E_x}{\sqrt{N}\sigma_x}\right)^2\right) dE \; ,$$

and having in mind that $\hat{\beta} = \beta N$ and $N \gg 1$ we finally obtain: $Z_N \sim \int_0^1 dx \int_{E_0}^{|E_0|} \exp\left(-N \cdot F(x,E)\right) dE$, where

$$F(x,E) = L(x) - \beta H(1-2x) + \beta E + \frac{1}{2N}\left(\frac{E-E_x}{\sigma_x}\right)^2 \; .$$

Taking into account Eqs. (2) and (3) we have

$$E_x = E_0(1-2x)^2, \; E_0 = E(\mathbf{s}_0) = -\frac{1}{2N}\sum_{ij} T_{ij}, \quad \sigma_x^2 = \frac{8\sum_{ij} T_{ij}^2}{N^2} x^2(1-x)^2 \; .$$

If the interaction between the nearest neighbors is $J$ it is easy to see that $E_0 = -DJ$ and $\sigma_x^2 = 16DJ^2 x^2(1-x)^2 / N$.

## Appendix B

Let us write down the system of equations (14) in details:

$$\frac{\ln\dfrac{1-x}{x}}{2(1-2x)} = qb(1-b) + qb^2(1-2x)^2$$

$$\frac{\dfrac{\ln\dfrac{1-x}{x}}{x} - \dfrac{1-2x}{2x(1-x)}}{(1-2x)^2} = -4qb^2(1-2x), \qquad x < 1/2 .$$

We multiply the second equation by $(1-2x)/2$ and add it to the first one. Next, we divide the second equation by $(1-2x)^2$ and transform it to a handy form. Then we obtain the new system of equations:

$$qb(1-b) = \varphi(x), \qquad \text{where } \varphi(x) = \frac{1}{2}\left( 3\frac{\ln\dfrac{1-x}{x}}{2(1-2x)} - \frac{1}{4x(1-x)} \right),$$

$$qb^2 = \xi(x), \quad \text{where } \xi(x) = \frac{1}{2(1-2x)^2}\left( \frac{1}{4x(1-x)} - \frac{\ln\dfrac{1-x}{x}}{2(1-2x)} \right).$$

(B1)

The functions $\varphi(x)$ and $\xi(x)$ are convenient when writing the final answer. It is easy to show that

$$\lim_{x\to 0}\varphi(x) = -\infty, \ \lim_{x\to 1/2}\varphi(x) = 1, \lim_{x\to 0}\xi(x) = +\infty, \ \lim_{x\to 1/2}\xi(x) = 1/3.$$

The graphs of the functions $\varphi(x)$ and $\xi(x)$ are presented in Fig. 11.

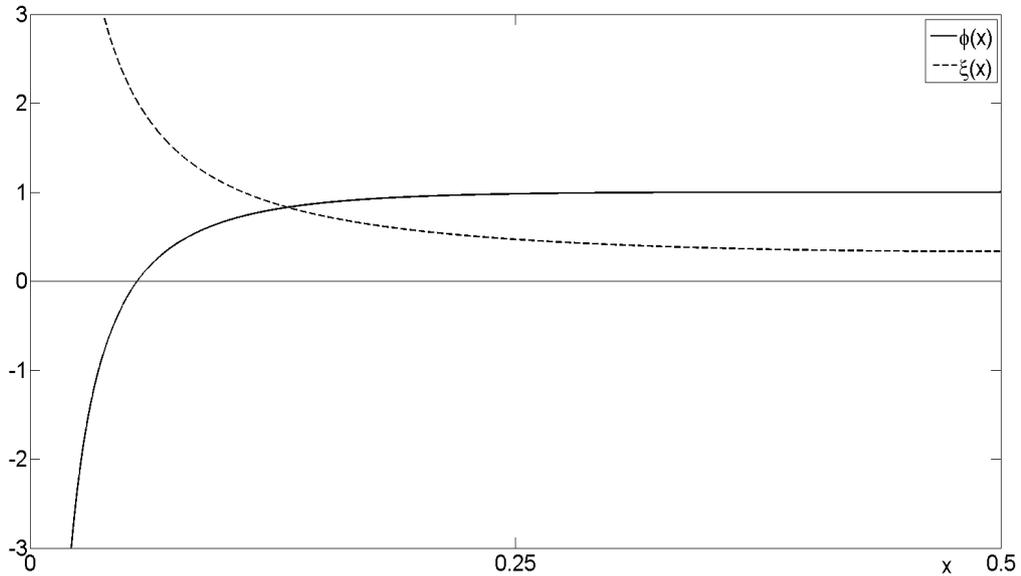

Fig. 11. The graphs of the auxiliary functions $\varphi(x)$ and $\xi(x)$.

Let us use the second equation (B1) to express $b$ in terms of $q$ and $x$. We substitute this expression in the left-hand side of the first of the equations (B1). The obtained equation contains $x$ and $q$ only:

$$\varphi(x) = \sqrt{q}\cdot\sqrt{\xi(x)} - \xi(x) .$$

Then inside the interval $(0, x_0)$ the coordinate of the tangency point of the curves has to satisfy the equation

$$G(x) = \frac{\left[\xi(x) + \varphi(x)\right]^2}{\xi(x)} = q \, . \tag{B2}$$

Substituting $q$ defined by Eq.(B2) into the second equation (B1) we obtain the value of $b$ at which the tangency of the curves $l(x)$ and $r(x,b)$ takes place:

$$b = B(x) = \frac{\xi(x)}{\xi(x) + \varphi(x)} \, .$$

The function $G(x)$ monotonically increases (see Fig.2). It is easy to calculate its limit at the right end of the interval $[0, 1/2]$: $\lim_{x \to 1/2} G(x) = 16/3$. The function $B(x)$ decreases monotonically at the interval $[0, 1/2]$. Its value at the right end of this interval is $\lim_{x \to 1/2} B(x) = 0.25$.

## Appendix C

Let us write down the explicit form of the system of equations (19):

$$\begin{aligned} \frac{\ln \dfrac{1-x}{x}}{2(1-2x)} &= qb(1-b) + qb^2(1-2x)^2, \\ L(x) - \frac{qb}{2}\left[(1-2x)^2 + 8b\left(x(1-x)\right)^2\right] &= -\ln 2 - \frac{qb^2}{4} \end{aligned} \qquad , \quad b \le 1 \, . \tag{C1}$$

Here $L(x)$ is given by Eq.(4). Rearranging the terms of the first equation (C1) and transforming the second equation with the aid of the line of equations:

$$L(x) + \ln 2 = \frac{qb}{4}(1-2x)^2\left[2 - b - 4bx(1-x)\right] = \frac{qb}{4}(1-2x)^2\left[1 - b + \frac{\ln \dfrac{1-x}{x}}{2(1-2x)}\right],$$

we obtain a new system of equations

$$qb^2 = \xi_1(x), \text{ where } \xi_1(x) = \frac{2}{(1-2x)^2}\left[\frac{\ln \dfrac{1-x}{x}}{2(1-2x)} - \frac{2\left(L(x) + \ln 2\right)}{(1-2x)^2}\right],$$

$$qb(1-b) = \varphi_1(x), \qquad \text{where} \quad \varphi_1(x) = 4\frac{L(x) + \ln 2}{(1-2x)^2} - \frac{\ln \dfrac{1-x}{x}}{2(1-2x)} . \tag{C2}$$

The functions $\varphi_1(x)$ and $\xi_1(x)$ are much like the functions $\varphi(x)$ and $\xi(x)$ defined in Eq.(B1); their graphs are shown in Fig. 12. It is easy to show that

$$\lim_{x \to 0} \varphi_1(x) = -\infty, \ \lim_{x \to 1/2} \varphi_1(x) = 1, \lim_{x \to 0} \xi_1(x) = +\infty, \ \lim_{x \to 1/2} \xi_1(x) = 1/3.$$

At the left end of the interval, when $x \to 0$, these functions tend to infinity as logarithms.

Expressing $b$ in terms of $q$ and $x$ with the aid of the first equation (C2) and substituting the result in the left-hand side of the second equation, we obtain

$$G_1(x) = \frac{\left[\xi_1(x) + \varphi_1(x)\right]^2}{\xi_1(x)} = q \, . \tag{C3}$$

The solution of this equation is the coordinate $x_j$ at which the depth of the local minimum becomes equal to the depth $f(x_0)$ of the minimum at the point $x_0$. The graph of the function $G_1(x)$ is shown in Fig. 5.

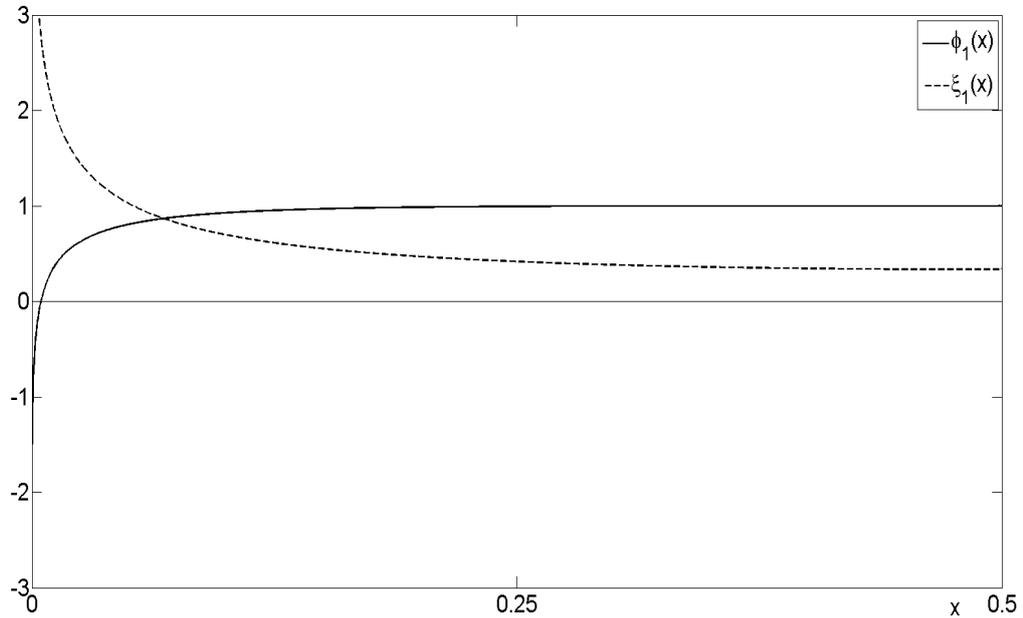

Fig. 12. The graphs of the functions $\varphi_1(x)$ and $\xi_1(x)$.

Substituting the left-hand side of (C3) in the first equation (C2) instead of $q$ we obtain the value of $b$ at which the both minima have the same depths:

$$b = B_1(x) = \frac{\xi_1(x)}{\xi_1(x) + \varphi_1(x)}.$$

The value of the parameter $b$ at which the global minimum jumps from the point $x_0$ to the point $x_j$ is $b_j = B_1(x_j)$. The graph of the function $B_1(x)$ is shown in Fig. 5.